\documentclass[11pt]{article}

\title{
Dilatons in curved backgrounds  by the Poisson--Lie transformation}

\def\ba{\begin{array}}
\def\ea{\end{array}}
\def\be{\begin{equation}}
\def\ee{\end{equation}}
\def\lbl{\label}
\def \rf{(\ref}

\def\x0{\x_0}
\def\x1{\x_1}

\def\eqn{equation}
\def\cond{condition}
\def\tfn{transformation}
\def\soln{solution}

\def\sm{$\sigma$--model}
\def\pl{Poisson--Lie}
\def\pltfn{Poisson--Lie transformation}
\def\dd{Drinfel'd double}
\def\vbe{vanishing $\beta$ equations}
\def\3dial{three--dimensional}
\def\-1{^{-1}}

\def\real{{\bf R}}

\def\cd{{\cal {D}}}
\def\cg{{\cal {G}}}
\def\tcg{{\tilde{\cal G}}}
\newtheorem{theorem}{Theorem}
\author{L. Hlavat\'y \\
Faculty of Nuclear Sciences and Physical Engineering,
\\ Czech Technical University,
\\ B\v rehov\'a 7, 115 19 Prague 1, Czech Republic
\\ 
{hlavaty@fjfi.cvut.cz} }

\begin{document}
\maketitle

\abstract{Transformations between group coordinates of three--dimensional conformal \sm s in the flat background
and their flat, i.e. Riemannian coordinates enable to find general dilaton fields for three--dimensional flat
\sm s. By the \pltfn {} we can get dilatons for the dual \sm s in a curved background. Unfortunately, in some
cases the dilatons depend on inadmissible auxiliary variables so the procedure is not universal. The cases where
the procedure gives proper and nontrivial dilatons in curved backgrounds are investigated and results given.}

\section{Introduction}\label{intro}
In the paper  \cite{hlasno:3dsm2} we have investigated conformally invariant \3dial \sm s on solvable Lie groups
that were \pl {} T--dual or plural to \sm s in the flat background. Several of them were nontrivial in the sense
that they lived in a curved background and had nonvanishing torsion. In some cases we were not able to find the
dilaton fields by the plurality procedure given in \cite{unge:pltp} because necessary conditions for application
of \pl {} \tfn {} were not satisfied for the constant dilaton.

Recently we have found explicit forms of transformations between the group coordinates of the flat \sm s  and
their flat coordinates, i.e. we expressed the Riemannian coordinates of the flat metric in parameters of its
solvable isometry subgroups \cite{hlatur:fc}. This enables us to write down the general form of the dilaton
field satisfying the \vbe {} for the flat model in terms of the group coordinates and consequently the dilaton
fields on the dual or plural nontrivial models.

To set our notation let us very briefly review the construction of the Poisson--Lie T--plural \sm s by means of
\dd s (For more detailed description see \cite{klse:dna}, \cite{kli:pltd}, \cite{unge:pltp},
\cite{hlasno:3dsm2}). The Lagrangian of dualizable \sm s can be written in terms of right--invariant fields on a
Lie group  $G$ that is a subgroup of the \dd {} as \be {L}=F_{ij}(\phi)\partial_- \phi^i\partial_+
\phi^j=E_{ab}(g)(\partial_- gg^{-1})^a(\partial_+ gg^{-1})^b, \lbl{rflag} \ee where\be \phi
:\real^2\rightarrow\, \real^n, \ \ \ F_{ij}(y)=e_i^a (g(y))E_{ab}(g(y)) e_j^b(g(y)),\lbl{metorze} \ee $e_i^a$
are components of right--invariant forms (vielbeins)
 $e_i^a(g)= \left( ({\rm d}g)_i \  . g^{-1}  \right)^a$ and $y^i$ are local coordinates of $g\in G$.
 \be
E(g)=(E_0^{-1}+\Pi(g))^{-1}, \ \ \ \Pi(g)=b(g)a(g)^{-1} = -\Pi(g)^t,\lbl{poiss}\ee and $a(g),b(g),d(g)$ are
submatrices of the adjoint representation of the group $G$ on the Lie algebra of the \dd {}
\footnote{ t denotes transposition.} \be Ad(g)^t  =  \left ( \begin{array}{cc}
  a(g)&0  \\ b(g)&d(g)  \end{array} \right ). \lbl{adg}\ee

The fact that for a \dd {} several decompositions of its Lie algebra $\cd$ into Manin triples {$(\cg|\tcg$)} may
exist leads to the notion of Poisson--Lie T--plurality \cite{unge:pltp}. Namely, let $\{ X_j,\tilde X^k\},\
j,k\in\{1,...,n\}$ {be} generators of Lie subalgebras {$\cg,\,\tcg$} of the Manin triple associated with the
Lagrangian \rf{rflag}) and $\{ U_j,\tilde U^k\}$ are generators of some other Manin triple {$(\cg_U|\tcg_U)$} in
the same \dd {} related by the $2n\times 2n$ transformation matrix as \be {\left (
\begin{array}{c}
  \vec X \\ \vec {\tilde X} \end{array} \right )}
  =  \left ( \begin{array}{cc}
 P&T  \\ R&S  \end{array} \right )\left ( \begin{array}{c} \vec U\\ \vec {\tilde  U} \end{array} \right),
\lbl{trsfmat}\ee where
$$ \vec X=(X_1, \ldots, X_n)^t, \ldots, \; \vec {\tilde  U} =(\tilde U^1,\ldots,\tilde U^n)^t.$$
The transformed model is then given by {the} Lagrangian of the form \rf{rflag}) but with $E(g)$ replaced by \be
E_U(g_u)=M(N+\Pi_U\,M)^{-1}=(\widetilde E_{0}^{-1}+\Pi_U)^{-1},\lbl{eg} \ee where
 \be M=S^tE_0-T^t, \ \
N=P^t-R^t E_0,\ \  \widetilde E_{0}=MN^{-1} \lbl{mn} \ee
 and $\Pi_U$ is calculated by \rf{poiss})
from the adjoint representation of the group $G_U$ generated by $\{U_j\}$. Note that for $P=S=0,\ T=R={\bf 1}$
we get the dual model with $\widetilde E_{0}=E_0^{-1}$, corresponding to the interchange $\cg \leftrightarrow
\tcg$ so that the duality transformation is a special case of the plurality transformation \rf{trsfmat}) --
\rf{mn}).

\section{\pl {} \tfn {} of  dilatons}
In quantum theory the duality or plurality transformation must be supplemented by a correction that comes from
integrating out the fields on the dual group $\tilde G$ in path integral formulation. In some cases it can be
absorbed at the 1-loop level into the transformation of the dilaton field $\Phi$ satisfying the so called \vbe
\begin{eqnarray}
\label{bt1} 0 & = & R_{ij}-\bigtriangledown_i\bigtriangledown_j\Phi- \frac{1}{4}H_{imn}H_j^{mn}, \\
 \label{bt2} 0 & = & H_{kij} \bigtriangledown^k\Phi+\bigtriangledown^k H_{kij},  \\
 \label{bt3} 0 & = & R-2\bigtriangledown_k\bigtriangledown^k\Phi- \bigtriangledown_k\Phi\bigtriangledown^k\Phi- \frac{1}{12}H_{kmn}H^{kmn}
 \end{eqnarray}
where the covariant derivatives $\bigtriangledown_k$, Ricci tensor $R_{ij}$ and Gauss curvature $R$ are
calculated from the metric \be G_{ij}=\frac{1}{2}(F_{ij}+F_{ji}) \ee that is also used for lowering and raising
indices, and the torsion is \be H_{ijk}=\partial_i B_{jk}+\partial_j B_{ki}+\partial_k B_{ij} \lbl{torsion} \ee
where \be B_{ij}=\frac{1}{2}(F_{ij}-F_{ji}). \lbl{torpot}\ee

The \pltfn {} of the tensor $F$ that follows from \rf{eg}) then must be accompanied by the transformation of the
dilaton \cite{unge:pltp} \be \Phi_U= \Phi+{\rm ln | Det} (N + \Pi_U M)| - {\rm ln| Det} ({\bf 1} + \Pi
E_0)|+{\rm ln | Det}\,a_U|-{\rm ln |  Det} \, a| \lbl{dualdil}\ee where $\Pi_U,a_U,$ are calculated by
\rf{poiss}) and \rf{adg}) but from the adjoint representation of the group $G_U$. The transformed dilaton
$\Phi_U$ then satisfy the \vbe {} if the dilaton $\Phi$ does.

Unfortunately, the right-hand side of the formula (\ref{dualdil}) may depend on the coordinates of the auxiliary
group $\widetilde G$. That's why the transformation of the dilaton field cannot be applied in general but only
if the following theorem holds \cite{hlasno:3dsm2}
\begin{theorem}\label{thm1}
The dilaton (\ref{dualdil}) for the model defined on the group $G_U$ exists if and only if
$$ \tilde U \Phi^{(0)} (g.\tilde g)=
\frac{ {\rm d}}{{\rm d} t} \Phi^{(0)} \left( g.\tilde g. \ {\rm exp}(t \tilde U) \right) |_{t=0} =0, \; \forall
g \in G_U, \; \forall \tilde g \in \tilde G_U, \; \forall \tilde U \in \tcg_U,   $$ where $\tilde U \in \tcg_U$
is extended as a left--invariant vector field on $D$ and \be\label{Phi0} \Phi^{(0)}(g) = \Phi(g) - {\rm ln |
Det} ({\bf 1} + \Pi(g) E_0)|-{\rm ln | Det} \, a(g)|. \ee
\end{theorem}
For applications it is much easier to check a weaker necessary condition.
\begin{theorem}\label{thm2}
A necessary condition for the existence of the dilaton (\ref{dualdil}) for the model defined on the group $G_U$
is \be \tilde U \Phi^{(0)} ( e ) = \frac{ {\rm d}}{{\rm d} t} \Phi^{(0)} ({\rm exp}(t \tilde U) )|_{t=0} =0, \;
\forall \tilde U \in \tcg_U,\label{cond2}\ee where $e$ is the unit of the \dd {} $D$.
\end{theorem}
For parametrization of $g\in G$ in the form \be g(y)=\exp (y_1X_1)\ \exp (y_2X_2)\ \exp (y_3X_3),\lbl{parg} \ee
where $y_j$ are coordinates {on} the group manifold and $X_j$ are the group generators the condition
(\ref{cond2}) can be rewritten (see \cite{hlasno:3dsm2}) as \be R^{jk}\frac{\partial \Phi^{(0)}(y)}{\partial
y_j}|_{y=0}=0,\label{checkthm2} \ee where $R$ is the submatrix in (\ref{trsfmat}).

The condition \rf{checkthm2}) could not be satisfied for some of the 
\sm s with constant dilaton field so that we were not able to find the transformed dilaton $\Phi_U$ that satisfy
the \vbe. The possibility to find the general dilaton fields for the flat models offers a possibility to
overcome this obstacle and obtain more general dilatons in curved backgrounds.
\section{Dilatons of \sm s on solvable three-dimensional groups}\lbl{3dsm}
All models investigated in the following admit nonsymmetric tensor $F_{ij}$ but their torsions vanish so without
loss of generality we shall deal with models having $F_{ij}=F_{ji}=G_{ij}$.

\subsection{General dilatons in flat backgrounds}
It follows from the construction of classical dualizable models that they are given by decompositions (Manin
triples) $(\cg|\tcg)$ and matrices $E_0$. Most of the flat and torsionless models found in \cite{hlasno:3dsm2}
can be formulated on the the \dd s with semiabelian decompositions $(X|{\bf 1})$ where {\bf 1} is the \3dial
abelian algebra. From the form of the \vbe {} (\ref{bt1}--\ref{bt3}) it is easy to see that the general form of
their dilaton fields is
\begin{equation}\label{gd}
    \Phi(y)=c_1\xi_{1}(y)+c_2\xi_{2}(y)+c_3\xi_{3}(y)+c_0,
\end{equation}where $\xi_{j}(y)$ are coordinates that bring  the flat metric $G_{ij}(y)$ to a constant form
$G'_{ij}$ (see \cite{hlatur:fc}) and $c_j$ are real constants satisfying
\be\sum_{j=1}^3G'^{ij}c_ic_j=0.\label{eqforc2}\ee

By the \pl {} \tfn {} of (\ref{gd}) we can get dilatons for the dual  \sm s but, as mentioned before, only if
the necessary conditions are satisfied.  Due to (\ref{Phi0}) and (\ref{gd}) the condition (\ref{checkthm2})
reads
\begin{equation}\label{cond2new}
    R^{jk}\left(c_m\frac{\partial \xi_m}{\partial y_j}-\frac{\partial}{\partial y_j} {\rm ln |\,
Det}[\,a(g)\left({\bf 1}+\Pi(g)E_0\right)\,]\,\,|\right)|_{y=0}=0.
\end{equation}
Moreover, the matrix $\Pi(g)$ vanishes for $ (X|{\bf1})$ and the flat coordinates can be chosen to satisfy
$\frac{\partial \xi_m}{\partial y_j}(0)=\delta_{mj}$. The condition (\ref{cond2new}) then simplifies to
\begin{equation}\label{eqforc1}
    R^{jk}\left(c_j-\frac{\partial}{\partial y_j} {\rm ln |\,
Det}\,a(g)\,|\right)|_{y=0}=0.
\end{equation}

\subsection{Dilatons for \sm s dual to ${\bf (5|1)}$}
The first \sm {} in the curved background we are going to investigate is given by the metric
\begin{equation}\label{g601}
    \widetilde G_{ij}(u)=\left(
\begin{array}{ccc}
 e^{-2\epsilon u_3} Q & \epsilon \,e^{-2\epsilon u_3} Q & V \cosh u_3-H \sinh u_3 \\
 \epsilon \,e^{-2\epsilon u_3} Q & e^{-2\epsilon u_3} Q & H \cosh u_3-V \sinh u_3  \\
 V \cosh u_3-H \sinh u_3 & H \cosh u_3-V \sinh u_3 & J
\end{array}
\right),
\end{equation}
where $\epsilon =\pm 1$ and $Q,V,H,J$ are constants. This metric has nonvanishing Ricci tensor but its Gauss
curvature is zero. It belongs to the \sm {} corresponding to the  ${(6_0|1)}$ decomposition of the $DD11$ (for
notation see \cite{snohla:ddoubles}) and $\widetilde E_0=\widetilde G(0)$. On the other hand, it can be obtained
by the \pltfn {} (\ref{eg}), (\ref{mn}) from the metric
\begin{equation}\label{g51b}
    G_{ij}(y)=\left(\matrix{ 0&0&v\,{\rm e}^{-y_1}
    \cr 0&q\,{\rm e}^{-2y_1}&0
    \cr v\,{\rm e}^{-y_1}&0&0
 \cr  } \right),\end{equation}
where $q,v$ are constants. The latter metric is flat and corresponds to the $(5|1)$ decomposition of the $DD11$
and $E_0=G(0)$.

The matrix (\ref{trsfmat}) that transform the Manin triple $(5|1)$ to $(6_0|1)$ and the metric (\ref{g51b}) to
(\ref{g601}) is
\begin{equation}\label{C60151}
 \left ( \begin{array}{cc}
 P&T  \\ R&S  \end{array} \right )=\left(
\begin{array}{cccccc}
 -\beta+\frac{1}{2} \alpha &\epsilon (\beta+\frac{1}{2} \alpha) & -\epsilon & 0 & 0 & 0 \\
 0 & 0 & 0 & \epsilon & 1 & \alpha \\
 -\epsilon & 1 & 0 & 0 & 0 & 0 \\
 0 & 0 & 0 & 0 & 0 & -\epsilon \\
 \frac{1}{2}\epsilon & \frac{1}{2} & 0 & 0 & 0 & 0 \\
 0 & 0 & 0 & -\frac{1}{2}\epsilon & \frac{1}{2} & \beta
\end{array}
\right),\ee where relations between the constants are \be \label{relconst}q=Q\-1,\ v=V-\epsilon\,H,\
\alpha=\frac{\epsilon\,V+H}{2\,Q},\ \beta= \epsilon\frac{\alpha^2 Q-J}{2v}.\ee In fact, the metric (\ref{g601})
is the most general that can be obtained by the \pltfn {} from a flat metric corresponding to the $(5|1)$
decomposition of the $DD11$.

General form of the dilaton field for the metric (\ref{g51b}) is given by (\ref{gd}) where
\begin{eqnarray}
\xi_{1}(y_{1},y_{2},y_{3})&=&-e^{-y_{1}}\nonumber\\
\xi_{2}(y_{1},y_{2},y_{3})&=&e^{-y_1} y_2\label{fc51}\\
\xi_{3}(y_{1},y_{2},y_{3})&=&\frac{q}{2 v}e^{-y_1}  {y_2}^2+y_3.\nonumber
\end{eqnarray} These are the coordinates that bring  the flat metric to its constant
form $G'_{ij}=G_{ij}(0)$ \cite{hlatur:fc}.

The formula \rf{dualdil}) for the general dilaton of the \sm {} given by (\ref{g601}) yields
\begin{equation}\label{dil601vy} \Phi_U(y)= -2 y_1+c_1e^{-y_1}+c_2e^{-y_1}y_2+c_3 \left(\frac{q}{2v}
   e^{-y_1} {y_2}^2+y_3\right) + c_0,
\end{equation}where the
coefficients satisfy the \eqn {} (\ref{eqforc2}) that in this case reads \begin{equation}\label{eqforc251}
 {v\,  c_2^2}+{2q}\,{c_1  c_3}=0.
\end{equation}
However, this is not yet the final form of the dilaton field because it is expressed in terms of the coordinates
$y$ of the \sm {} given by (\ref{g51b}) and it must be transformed to the coordinates $u$ of the \sm {} given by
(\ref{g601}). The transformation formulas between these coordinates follow from two different decompositions of
elements of the \dd {} $DD11$, namely from the relation \be e^{-y_1X_1}e^{-y_2X_2}e^{-y_3X_3}e^{-\tilde
y_1\widetilde X_1}e^{-\tilde y_2\widetilde X_2}e^{-\tilde y_3\widetilde
X_3}=e^{-u_3U_3}e^{-u_2U_2}e^{-u_1U_1}e^{-\tilde u_1\widetilde U_1}e^{-\tilde u_2\tilde U_2}e^{-\tilde
u_3\widetilde U_3}, \label{lgh}\ee where $X_j,\widetilde X_j$ are generators corresponding to the decomposition
$(5|1)$ of the \dd {} $DD11$ and  $U_j,\widetilde U_j$ are generators of the decomposition $(6_0|1)$. They can
be related by (\ref{C60151}). Coordinates $y$ in terms of $u$ are then expressed as
\begin{eqnarray}
  y_1 &=& -\epsilon\, u_3 \nonumber,\\
   y_2 &=& \frac{\epsilon\,\widetilde u_1+\widetilde u_2}{2}, \nonumber\\
   y_3 &=& \frac{-\epsilon\,u_1+u_2}{2}+\beta u_3, \\
   \widetilde y_1&=&\beta(-\widetilde u_1+\epsilon\,\widetilde u_2)-\epsilon\,\widetilde u_3 +
   \frac{1}{2}(\widetilde u_1+\epsilon\,\widetilde u_2)(\alpha+u_1+\epsilon\,u_2+\epsilon\,\alpha u_3)
   ,\nonumber\\
   \widetilde y_2 &=&\epsilon\,u_1+u_2+\alpha\,u_3, \nonumber\\
   \widetilde y_3 &=& -\epsilon\,\widetilde u_1+\widetilde u_2. \nonumber
\end{eqnarray}
We can see that unless $c_2=0,\,c_3=0$ the dilaton \rf{dil601vy}) depends on the coordinate
${\epsilon\,\widetilde u_1+\widetilde u_2}$. It is not admissible and thus the general form of dilaton obtained
by the \pltfn {} for the metric (\ref{g601}) is
\begin{equation}\label{dil601} \widetilde \Phi(u)= \Phi_U(y(u))=  2 \epsilon u_3+c_1 e^{{\epsilon }{u_3}} + c_0.
\end{equation}
We have checked that the \vbe {} for $\widetilde \Phi(u)$ and $\widetilde G_{ij}(u)$ given by \rf{g601}) are
satisfied.

Note that the condition $c_3=0, c_2=0$ is more strict than the necessary \cond {} (\ref{eqforc1}) that implies
$c_2=0$ only. It means that the necessary \cond {} (\ref{cond2}) is not sufficient for the \pltfn {} of the
dilaton.

By other plurality transformations of (\ref{g51b}) we can get \sm s with curved background corresponding to {the
decompositions} $(1|6_0)$ and  $(5.ii|6_0)$ of the DD11.  For the dilaton fields the formula \rf{dualdil}) could
be again used but we were not able to express the coordinates $y,\tilde y$ in terms of $u,\tilde u$ from the
relation (\ref{lgh}) in these cases.

\subsection{Sigma models dual to ${\bf (4|1)}$}
A bit more complicated \sm {} is given by the metric $\widetilde G_{ij}(u)$, where
\begin{eqnarray}\label{g602}
    \widetilde G_{11}(u)&=& \widetilde G_{22}(u)= e^{-2\epsilon u_3} Q\nonumber\\
    \widetilde G_{12}(u)&=& \widetilde G_{21}(u)= \epsilon\,e^{-2\epsilon u_3} Q\nonumber\\
    \widetilde G_{13}(u)&=& \widetilde G_{31}(u)= V \cosh u_3-H \sinh u_3 +Q\frac{\epsilon\,V-H}{2} u_3  e^{-u_3}\\
    \widetilde G_{23}(u)&=& \widetilde G_{32}(u)= H \cosh u_3-V \sinh u_3 +Q\frac{V-\epsilon\,H}{2} u_3  e^{-u_3} \nonumber\\
    \widetilde G_{33}(u)&=& J-Q\,({V-\epsilon\,H})^{2}u_3^2\nonumber
\end{eqnarray}
where $\epsilon =\pm 1$ and $Q,V,H,J$ are constants. Again, this metric has nonvanishing Ricci tensor and its
Gauss curvature is zero. It belongs to the \sm {} corresponding to the  $(6_0|2)$ decomposition of the $DD12$
and $\widetilde E_0=\widetilde G(0)$. Besides that it can be obtained by the \pltfn {} (\ref{eg}), (\ref{mn})
from the metric \begin{equation}\label{g41b}
    G_{ij}(y)=\left(\matrix{ 0&v\,{\rm e}^{-y_1}y_1&v\,{\rm e}^{-y_1}
    \cr v\,{\rm e}^{-y_1}y_1&q\,{\rm e}^{-2y_1}&0
    \cr v\,{\rm e}^{-y_1}&0&0
 \cr  } \right),\end{equation}
where $q,v$ are constants. This metric is flat and corresponds to the $(4|1)$ decomposition of the $DD12$.

The matrix (\ref{trsfmat}) that transform the metric (\ref{g41b}) to (\ref{g602}) is
\begin{equation}
\label{pqrs41}\left (
\begin{array}{cc}
 P&T  \\ R&S  \end{array} \right )=\left(
\begin{array}{cccccc}
 -\beta-\frac{1}{2} \alpha &\epsilon (\beta-\frac{1}{2} \alpha) & -\epsilon & 0 & 0 & 0 \\
 0 & 0 & 0 & -\epsilon & -1 & \alpha \\
 -\epsilon & 1 & 0 & 0 & 0 & 0 \\
 0 & 0 & 0 & 0 & 0 & -\epsilon \\
 -\frac{1}{2}\epsilon & -\frac{1}{2} & 0 & 0 & 0 & 0 \\
 0 & 0 & 0 & -\frac{1}{2}\epsilon & \frac{1}{2} & \beta
\end{array}
\right),\ee where the relations between the constants are \be \label{relconst2}q=Q\-1,\ v=V-\epsilon\,H,\
\alpha=-\frac{\epsilon\,V+H}{2\,Q},\ \beta= \epsilon\frac{\alpha^2 Q-J}{2v}.\ee

The dilaton field for the \sm {} given by (\ref{g602}) obtained by insertion of the flat coordinates of the
metric (\ref{g41b}) into the formula \rf{dualdil}) is
\begin{eqnarray}\label{dil602vy} \Phi(y)&=& -2 y_1+ c_1e^{-y_1}+c_2 \left(\frac{v}{q}
   \left(y_1+e^{-y_1}\right)+e^{-y_1} y_2\right)+\nonumber\\
   & & c_3
   \left(\frac{v}{q}(y_1 -\sinh y_1)+
  \frac{q}{2v} e^{-y_1} y_2^2+\left(e^{-y_1}+ y_1 -1\right)
   y_2+y_3\right)+ c_0,
\end{eqnarray}where the
coefficients satisfy the \eqn {} (\ref{eqforc251}). To get the final form of the dilaton field we must transform
it to the coordinates $u$. The transformation formulas follow from decompositions of elements of the \dd {}
$DD12$, namely from the relation (\ref{lgh}) where $X_j,\widetilde X_j$ are generators corresponding to the
decomposition $(4|1)$ and  $U_j,\widetilde U_j$, related by \rf{trsfmat}) and (\ref{pqrs41}), correspond to the
decomposition $(6_0|2)$ . Coordinates $y$ in terms of $u$ are then expressed as \begin{eqnarray}
  y_1 &=& -\epsilon\, u_3 \nonumber,\\
   y_2 &=& -\frac{\epsilon\,\widetilde u_1+\widetilde u_2}{2}, \nonumber\\
   y_3 &=& \frac{-\epsilon\,u_1+u_2}{2}+\beta u_3, \\
   \widetilde y_1&=&\beta(-\widetilde u_1+\epsilon\,\widetilde u_2)+\frac{1}{2}(\widetilde u_1+\epsilon\,\widetilde u_2)
   (-\alpha+u_1+\epsilon\,u_2 -\epsilon\,\alpha\,u_3)
   -\frac{1}{4}{\widetilde u_1}^2+\frac{1}{2}\epsilon\,\widetilde u_1\widetilde u_2+\frac{1}{4}{\widetilde u_2}^2
   -\epsilon\,\widetilde u_3   ,\nonumber\\
   \widetilde y_2 &=&-\epsilon\,u_1-u_2+\alpha\,u_3, \nonumber\\
   \widetilde y_3 &=& -\epsilon\,\widetilde u_1+\widetilde u_2. \nonumber
\end{eqnarray}
In order that the dilaton does not depend on the coordinate ${\epsilon\,\widetilde u_1+\widetilde u_2}$ we must
set $c_2=0,\,c_3=0$ and the general form of the dilaton obtained by the \pltfn {} for the metric (\ref{g602}) is
again (\ref{dil601}). The \vbe {} are satisfied.

Let us mention in the end that there are still other models with curved backgrounds dual to the flat ones,
namely those corresponding to the Manin triples  $(1|2),\, (1|7_0)$ of the \dd s DD15 and DD19.
Unfortunately, in these cases all $y$ coordinates depend on the $\widetilde u_1,\widetilde u_2,\widetilde u_3$
so that only $\Phi=const$ may be inserted into \rf{dualdil}) giving results published in \cite{hlasno:3dsm2}.

\section{Conclusions}
We have investigated the possibilities to apply the \pltfn {} to the general solution of the \vbe {} for the
flat metric.
We have obtained dilaton fields for the metrics (\ref{g601}) and (\ref{g602}) having a nontrivial Ricci tensor.
They are the most general dilatons that can be obtained by the \pltfn {} from the general dilatons
\rf{dil601vy}), \rf{dil602vy}) of the dual flat metrics (\ref{g51b}) and (\ref{g41b}). An interesting but yet
unsolved question is whether the dilaton \rf{dil601}) is the general \soln {} of the \vbe{} for the curved
backgrounds (\ref{g601}) and (\ref{g602}).

On the other hand, we have found that the procedure does not work universally because the transformed dilatons
often depend on inadmissible auxiliary variables and the above mentioned cases show that the necessary condition
\rf{cond2}) for the applicability of the formula (\ref{dualdil}) is not sufficient.
\section{Acknowledgements}
This work was supported by the project of the Grant Agency of the Czech Republic No. 202/06/1480 and by the
research plan LC527 15397/2005--31 of the Ministry of Education of the Czech Republic. Useful comments of Libor
\v Snobl are gratefully acknowledged.

\end{document}